\documentclass[conference,12pt]{IEEEtran}
\usepackage{epsfig}
\usepackage{graphicx}
\usepackage{cite}
\usepackage{amsmath}
\usepackage{amssymb}

\newcommand{\ignore}[1]{}
\newcommand{\comment}[1]  {}
\DeclareMathOperator*{\argmax}{arg\,max}
\newcommand\ie{{\textsl{i.e.\,}}}

\newcommand\etal{{\textsl{et al.\,}}}
\def\BE{\begin{equation}}
\def\EE{\end{equation}}
\def\BEA{\begin{eqnarray}}
\def\EEA{\end{eqnarray}}
\newcommand{\cut}[1]{{}}
\newcommand\vb{{\bf b}}
\newcommand\vc{{\bf c}}

\newcommand\vg{{\bf g}}

\newcommand\vx{{\bf x}}
\newcommand\vy{{\bf y}}
\newcommand\vz{{\bf z}}
\newcommand\mA{{\bf A}} %matrix

\newcommand\mC{{\bf C}}

\newcommand\mF{{\bf F}}

\newcommand\mI{{\bf I}}

\newcommand\mV{{\bf V}}

\newcommand\mX{{\bf X}}

\newcommand\mZ{{\bf Z}}
\def\ceil#1{\lceil #1 \rceil}

\begin{document}
\twocolumn
%\onecolumn
\title{Polynomial Linear Programming\\with Gaussian Belief Propagation}


\author{\IEEEauthorblockN{Danny Bickson, Yoav Tock}
\IEEEauthorblockA{IBM Haifa Research Lab\\
Mount Carmel\\Haifa 31905, Israel\\
Email: \{dannybi,tock\}@il.ibm.com\\} \and \IEEEauthorblockN{Ori
Shental} \IEEEauthorblockA{Center for Magnetic\\Recording
Research\\UCSD, San Diego\\9500 Gilman
Drive\\La Jolla, CA 92093, USA\\Email: oshental@ucsd.edu} \and
\IEEEauthorblockN{Danny Dolev} \IEEEauthorblockA{School of
Computer Science\\and Engineering\\Hebrew University of
Jerusalem\\Jerusalem 91904, Israel\\Email: dolev@cs.huji.ac.il}}


%Telephone: (800) 555--1212, Fax: (888) 555--1212}


\maketitle

\begin{abstract}
%\boldmath
Interior-point methods are state-of-the-art algorithms for solving
linear programming (LP) problems with polynomial complexity.
Specifically, the Karmarkar algorithm typically solves LP problems
in time $O(n^{3.5})$, where $n$ is the number of unknown
variables. Karmarkar's celebrated algorithm is known to be an
instance of the log-barrier method using the Newton iteration. The
main computational overhead of this method is in inverting the
Hessian matrix of the Newton iteration. In this contribution, we
propose the application of the  Gaussian belief propagation (GaBP)
algorithm as part of an efficient and distributed LP solver that
exploits the sparse and symmetric structure of the Hessian matrix
and avoids the need for direct matrix inversion. This approach
shifts the computation from realm of linear algebra to that of
probabilistic inference on graphical models, thus applying GaBP as
an efficient inference engine. Our construction is general and can
be used for any interior-point algorithm which uses the Newton
method, including non-linear program solvers.

\end{abstract}


% \fi
\IEEEpeerreviewmaketitle

\section{Introduction}
In recent years, considerable attention has been dedicated to the
relation between belief propagation message passing and linear
programming schemes. This relation is natural since the maximum
a-posteriori (MAP) inference problem can be translated into
integer linear programming (ILP)~\cite{LPWeiss}.

Weiss \etal~\cite{LPWeiss} approximate the solution to the ILP
problem by relaxing it to a LP problem using convex variational
methods. In~\cite{LPWeiss2}, tree-reweighted belief propagation
(BP) is used to find the global minimum of a convex approximation
to the free energy. Both of these works apply discrete forms of
BP. Globerson \etal~\cite{Globerson,Globerson2} assume convexity
of the problem and modify the BP update rules using
dual-coordinate ascent algorithm. Hazan \etal~\cite{Hazan}
describe an algorithm for solving a general convex free energy
minimization. In both cases the algorithm is guaranteed to
converge to the global minimum as the problem is tailored to be
convex.

In the present work we take a different path. Unlike most of the
previous work which uses gradient-descent methods, we show how to
use interior-point methods which are shown to have strong
advantages over gradient and steepest descent methods. (For a
comparative study see \cite[$\S 9.5$,p. 496]{BV04}.) The main
benefit of using interior point methods is their rapid
convergence, which is quadratic once we are close enough to the
optimal solution. Their main drawback is that they require heavier
computational effort for forming and inverting the Hessian matrix,
needed for computing the Newton step. To overcome this, we propose
the use of Gaussian BP (GaBP)~\cite{ISIT1,Allerton}, which is a
variant of BP applicable when the underlying distribution is
Gaussian. Using GaBP, we are able to reduce the time associated
with the Hessian inversion task, from $O(n^{2.5})$ to
$O(np{\log(\epsilon)}/{\log(\gamma)})$ at the worst case, where
$p<n$ is the size of the constraint matrix $\mA$, $\epsilon$ is
the desired accuracy, and $1/2 < \gamma < 1$ is a parameter
characterizing the matrix $\mA$. This computational saving is
accomplished by exploiting the sparsity of the Hessian matrix.

An additional benefit of our GaBP-based approach is that the
polynomial-complexity LP solver can be implemented in a
distributed manner, enabling efficient solution of large-scale
problems.

We also provide what we believe is the first theoretical analysis
of the convergence speed of the GaBP algorithm.

The paper is organized as follows. In Section
\ref{linear-programming}, we reduce standard linear programming to
a least-squares problem. Section \ref{lp-to-gm} shows how to solve
the least-squares problem using the GaBP algorithm. In Section
\ref{primal-dual}, we extend our construction to the primal-dual
method. We give our convergence results for the GaBP algorithm in
Section \ref{convergence}, and demonstrate our construction in
Section \ref{exp-results} using an elementary example. We present
our conclusions in Section~\ref{conc}.

\section{Standard Linear Programming}
\label{linear-programming} Consider the standard linear program
%\vspace{-2mm}
\begin{subequations}
\begin{eqnarray}
 \mbox{minimize}_{\vx} &  {\vc^T\vx } \label{can-linear} \\
\mbox{subject to} & \mA\vx = \vb,\ \ \ \ \vx\ge 0
\end{eqnarray}
\end{subequations}

  where $\mA \in \mathbb{R}^{n \times p}$ with
$\mathbf{rank}\{\mA\} = p < n$. We assume the problem is solvable
with an optimal $\vx^*$ assignment. We also assume that the
problem is strictly feasible, or in other words there exists $\vx
\in \mathbb{R}^n$ that satisfies $\mA\vx = \vb$ and $\vx > 0$.

Using the log-barrier method \cite[$\S 11.2$]{BV04}, one gets
%\vspace{-2mm}
\begin{subequations}
\begin{eqnarray}
 \mbox{minimize}_{\vx, \mu} & \vc^T\vx - \mu \Sigma_{k=1}^{n}
\log x_k \label{log-barrier}\\
 \mbox{subject to} & \mA\vx = \vb.
 \label{log-barrier-constraints}
\end{eqnarray}
\end{subequations}

 This is an approximation to the original problem
 (\ref{can-linear}). The quality of the approximation improves as
 the parameter $\mu \rightarrow 0$.

\begin{table*}[htb!]
\begin{center}
\normalsize \caption{The Newton algorithm \cite[$\S 9.5.2$]{BV04}
.}
\begin{tabular}{lcl}
  \hline
  Given & &  feasible starting point $\vx_0$ and tolerance $\epsilon > 0$, $k = 1$  \\
  \hline
  Repeat & 1 & Compute the Newton step and decrement \\
  & & $\Delta \vx = f''(\vx)^{-1} f'(\vx), \ \ \ \ \ \lambda^2 = f'(\vx)^T
  \Delta \vx$ \\
   & 2 & Stopping criterion. quit if $ \lambda^2/2 \le \epsilon$\\
   & 3 & Line search. Choose step size t by
backtracking line search. \\
  & 4 & Update. $\vx_k := \vx_{k-1} + t\Delta \vx, \ \ \ k = k + 1$ \\
  \hline
\label{newton-method}
\end{tabular}
\end{center}
\end{table*}

%\vspace{-8mm}
Now we would like to use the Newton method in for
solving the log-barrier constrained objective
function~(\ref{log-barrier}), described in Table
\ref{newton-method}. Suppose that we have an initial feasible
point $\vx_0$ for the canonical linear program~(\ref{can-linear}).
We approximate the objective function~(\ref{log-barrier}) around
the current point $\tilde{\vx}$ using a second-order Taylor
expansion \BE \label{taylor} f(\tilde{\vx} + \Delta \vx) \simeq
f(\tilde{\vx}) + f'(\tilde{\vx}) \Delta \vx + 1/2 \Delta \vx^T
f''(\tilde{\vx}) \Delta \vx. \EE Finding the optimal search
direction $\Delta \vx$ yields the computation of the gradient and
compare it to zero \BE \frac{\partial f}{\partial \Delta \vx} =
f'(\tilde{\vx}) + f''(\tilde{\vx}) \Delta \vx = 0,
\label{taylor-gradient} \EE \BE \Delta \vx = -
f''(\tilde{\vx})^{-1} f'(\tilde{\vx}). \label{newton-step} \EE

Denoting the current point $\tilde{\vx}\triangleq(\vx, \mu, \vy)$
and the Newton step $\Delta \vx \triangleq (\vx, \vy, \mu)$, we
compute the gradient
%TWOCOLS
\begin{subequations}
\begin{eqnarray*}
 f'(\vx,\mu,\vy) \equiv ({\partial f(\vx,
\mu, \vy) }/{ \partial \vx}, {\partial f(\vx, \mu, \vy) }/{
\partial \mu},\\
,{\partial f(\vx, \mu, \vy) }/{ \partial \vy})
\end{eqnarray*}
\end{subequations}
%ONECOL
%\begin{subequations}
%\begin{eqnarray*}
% f'(\vx,\mu,\vy) \equiv ({\partial f(\vx,
%\mu, \vy) }/{ \partial \vx}, {\partial f(\vx, \mu, \vy) }/{
%\partial \mu},{\partial f(\vx, \mu, \vy) }/{ \partial \vy})
%\end{eqnarray*}
%\end{subequations}

The Lagrangian is \BE \mathcal{L}( \vx, \mu, \vy) = \vc^T\vx  -
\mu \Sigma_k \log x_k + \vy^T(\vb - \mA\vx), \EE

\BE \frac{
\partial \mathcal{L}(\vx,\mu,\vy)}{
\partial \vx} = \vc - \mu \mX^{-1} \mathbf{1} - \vy^T\mA = 0
\label{log-bar-der-x-1}, \EE \BE \frac{
\partial^2 \mathcal{L}(\vx,\mu,\vy)}{
\partial \vx}= \mu \mX^{-2} \label{log-bar-der-x-2}, \EE where $\mX\triangleq\textrm{diag}(\vx)$ and $\mathbf{1}$ is the all-one column vector.
Substituting~(\ref{log-bar-der-x-1})-(\ref{log-bar-der-x-2}) into
(\ref{taylor-gradient}), we get \BE \vc - \mu \mX^{-1} \mathbf{1}
- \vy^T\mA + \mu \mX^{-2} \vx = 0, \EE \BE \vc - \mu \mX^{-1}
\mathbf{1}  + \vx \mu \mX^{-2} = \vy^T \mA
\label{first-order-condition1}, \EE

\BE \frac{
\partial \mathcal{L}(\vx,\mu,\vy)}{ \partial \vy } = \mA\vx = 0\label{log-bar-der-y-1}. \EE
Now multiplying (\ref{first-order-condition1}) by $\mA\mX^2$, and
using (\ref{log-bar-der-y-1}) to eliminate $\vx$ we get \BE
\mA\mX^2\mA^T\vy = \mA\mX^2\vc - \mu\mA\mX \mathbf{1}. \EE These
normal equations can be recognized as generated from the linear
least-squares problem \BE \min_\vy || \mX\mA^T\vy - \mX\vc -
\mu\mA\mX \mathbf{1} ||_2^2 \label{primal-ls}. \EE Solving for
$\vy$ we can compute the Newton direction $\vx$, taking a step
towards the boundary and compose one iteration of the Newton
algorithm. Next, we will explain how to shift the deterministic LP
problem to the probabilistic domain and solve it distributively
using GaBP.

\section{From LP to Probabilistic Inference} \label{lp-to-gm}
We start from the
least-squares problem~(\ref{primal-ls}), changing notations to \BE
\min_{\vy} ||\mF \vy - \vg ||^2_2, \label{ls-fg} \EE where $\mF
\triangleq \mX\mA^T, \vg \triangleq \mX\vc  + \mu \mA\mX
\mathbf{1}$. Now we define a multivariate Gaussian \BE
p(\hat{\vx}) \triangleq p(\vx,\vy) \propto \exp(-1/2(\mF \vy -
\vg)^T\mI(\mF \vy- \vg)). \label{gaussian-mult} \EE It is clear
that $\hat{\vy}$, the minimizing solution of (\ref{ls-fg}), is the
MAP estimator of the conditional probability

\BE \label{pi} \hat{\vy} = \argmax_{\vy} p(\vy|\vx) = \nonumber \EE
\BE = \mathcal{N}((\mF^T\mF)^{-1} \mF^T \vg , (\mF^T\mF)^{-1}). \EE

%\BE \hat{\vy} =
%\argmax_{\vy} p(\vy|\vx) =
%\mathcal{N}((\mF^T\mF)^{-1} \mF^T \vg , (\mF^T\mF)^{-1}). \label{pi} \EE

Recent results by Bickson and Shental~\etal
\cite{Allerton,ISIT1,ISIT2} show that the pseudoinverse
problem~(\ref{pi}) can be computed efficiently and distributively
by using the GaBP algorithm.

\ignore{ Schur complement of the following covariance matrix \BE
\mC = \left(%
\begin{array}{cc}
  -\mI & \mF \\
  \mF^T & \mathbf{0} \\
\end{array}%
\right) \label{C_cov} \EE }%ignore

The formulation~(\ref{gaussian-mult}) allows us to shift the
least-squares problem from an algebraic to a probabilistic domain.
Instead of solving a deterministic vector-matrix linear equation,
we now solve an inference problem in a graphical model describing
a certain Gaussian distribution function. Following \cite{ISIT2}
we define the joint covariance matrix \BE \mC \triangleq \left(%
\begin{array}{cc}
  -\mI & \mF \\
  \mF^T & \mathbf{0} \\
\end{array}%
\right) \label{C_cov} \EE and the shift vector $\vb
\triangleq\{\mathbf{0}^{T},\vg^{T}\}^{T}\in\mathbb{R}^{(p+n)\times
1}.$

Given the covariance matrix $\mC$ and the shift vector $\vb$, one
can write explicitly the Gaussian density function, $p(\hat{\vx})$
, and its corresponding graph $\mathcal{G}$ with edge potentials
(`compatibility functions') $\psi_{ij}$ and self-potentials
(`evidence') $\phi_{i}$. These graph potentials are determined
according to the following pairwise factorization of the Gaussian
distribution $p(\vx) \propto
\prod_{i=1}^{n}\phi_{i}(x_{i})\prod_{\{i,j\}}\psi_{ij}(x_{i},x_{j}),$
        resulting in $ \psi_{ij}(x_{i},x_{j})\triangleq \exp(-x_{i}C_{ij}x_{j}),$ and
        $ \phi_{i}(x_{i}) \triangleq
        \exp\big(b_{i}x_{i}-C_{ii}x_{i}^{2}/2\big).$
        The set of edges $\{i,j\}$ corresponds to the set of
        non-zero entries in $\mC$ (\ref{C_cov}). Hence, we would like to calculate the
marginal densities, which must also be Gaussian,
\BE p(x_{i})\sim\mathcal{N}(\mu_{i}=\{\mC^{-1}\vg\}_{i},P_{i}^{-1}=\{\mC^{-1}\}_{ii}),
\nonumber \EE
\BE \forall i > p, \nonumber \EE where $\mu_{i}$ and $P_{i}$ are the
marginal mean and inverse variance (a.k.a. precision),
respectively. Recall that, according to \cite{ISIT2}, the inferred
mean $\mu_{i}$ is identical to the desired solution $\hat{y}$ of
(\ref{pi}). \comment{The move to the probabilistic domain calls
for the utilization of BP as an efficient inference engine. GaBP
is a special case of continuous BP where the underlying
distribution is Gaussian. In~\cite{Allerton,ISIT1} we show how to
derive the GaBP update rules by substituting Gaussian
distributions in the continuous BP equations. The output of this
derivation is update rules that are computed locally by each
node.} The GaBP update rules are summarized in
Table~\ref{tab_summary}.

\ignore{ The sum-product rule of BP for \emph{continuous}
variables, required in our case, is given by~\cite{Weiss}
\begin{equation}\label{eq_contBP}
    m_{ij}(x_j) = \alpha \int_{x_i} \psi_{ij}(x_i,x_j) \phi_{i}(x_i)
\prod_{k \in \mathcal{N}(i)\setminus j} m_{ki}(x_i) dx_{i},
\end{equation} where $m_{ij}(x_j)$ is the message sent from node $i$ to node $j$ over their shared edge on the graph, scalar $\alpha$  is a normalization constant and the set $\mathcal{N}(i)\backslash j$ denotes all the nodes neighboring
$x_{i}$, except $x_{j}$. The marginals are computed according to
the product rule~\cite{Weiss} \BE\label{eq_product}
p(x_{i})=\alpha
\phi_{i}(x_{i})\prod_{k\in\mathcal{N}(i)}m_{ki}(x_{i}). \EE }

\begin{table*}[htb!]
\normalsize \caption{Computing $\vx = \mA^{-1}\vb$ via GaBP
\cite{ISIT1}.} \centerline{ \label{tab_summary}
\begin{tabular}{|c|c|l|}
  \hline
  \textbf{\#} & \textbf{Stage} & \textbf{Operation}\\
  \hline
  1. & \emph{Initialize} & Compute $P_{ii}=A_{ii}$ and $\mu_{ii}=b_{i}/A_{ii}$.\\
  && Set $P_{ki}=0$ and $\mu_{ki}=0$, $\forall k\neq i$.\\ \hline
  2. & \emph{Iterate} & Propagate $P_{ki}$ and $\mu_{ki}$, $\forall k\neq i \;
\mbox{\rm such that} \; A_{ki}\neq0$.\\& & Compute $P_{i\backslash
j}=P_{ii}+\sum_{{k}\in\mathbb{N}(i) \backslash j} P_{ki}$ and
$\mu_{i\backslash j} = P_{i\backslash
j}^{-1}(P_{ii}\mu_{ii}+\sum_{k \in \mathrm{N}(i) \backslash j}
P_{ki}\mu_{ki})$.\\
  && Compute $P_{ij} = -A_{ij}P_{i\backslash j}^{-1}A_{ji}$ and $\mu_{ij} =
-P_{ij}^{-1}A_{ij}\mu_{i\backslash j}$.\\\hline
  3. & \emph{Check} & If $P_{ij}$ and $\mu_{ij}$ did not converge, return to
    \#2. Else, continue to \#4.\\\hline
  4. & \emph{Infer} & $P_{i}=P_{ii}+\sum_{{k}\in\mathrm{N}(i)}
P_{ki}$ , $\mu_{i}=P_{i}^{-1}(P_{ii}\mu_{ii}+\sum_{k \in
\mathrm{N}(i)} P_{ki}\mu_{ki})$.\\
  \hline
  5. & \emph{Output} & $x_{i}= \mu_{i} $ \\\hline
\end{tabular}} %\vspace{0.5cm}
\end{table*}

It is known that if GaBP converges, it results in exact inference
~\cite{Weiss}. However,  in contrast to  conventional iterative
methods for the solution of systems of linear equations, for GaBP,
determining the exact region of convergence and convergence rate
remain open research problems. All that is known is a sufficient
(but not necessary) condition~\cite{WS,JMLR}
%stating that GaBP converges for the spectral
%radius\footnote{$\rho(\mB)\triangleq\max_{1\leq i\leq
stating that GaBP converges when the spectral radius satisfies
\mbox{$\rho(|\mI_{K}-\mA|)<1$}. A stricter sufficient
condition~\cite{Weiss}, determines that the matrix $\mA$ must be
diagonally dominant (\ie, $|a_{ii}|>\sum_{j\neq i}|a_{ij}| ,
\forall i$) in order for GaBP to converge. Convergence speed is
discussed in Section \ref{conv-speed}.

\section{Extending the Construction to the Primal-Dual Method}
\label{primal-dual} In the previous section we have shown how to
compute one iteration of the Newton method using GaBP. In this
section we extend the technique for computing the primal-dual
method. This construction is attractive, since the extended
technique has the same computation overhead.

The dual problem (\cite{GHare}) conforming to (\ref{can-linear})
can be computed using the Lagrangian \BE \mathcal{L}(\vx,\vy,\vz)
= \vc^T\vx + \vy^T(\vb - \mA\vx) - \vz^T\vx , \ \ \  \vz \ge 0,
\nonumber \EE
\begin{subequations}
\begin{eqnarray}
\label{lagrange1}
  g(\vy, \vz) = \inf_{x} \mathcal{\mathcal{L}}(\vx, \vy,
\vz), \\
  \mbox{subject to} \ \ \ \ \mA\vx = \vb, \vx \ge 0.
\end{eqnarray}
\end{subequations}
while
 \BE \frac{\partial \mathcal{L}(\vx,\vy,\vz) }{\partial \vx} =
\vc - \mA^T\vy - \vz = 0 \label{lagrange2}. \EE

Substituting (\ref{lagrange2}) into (\ref{lagrange1}) we get
\begin{subequations}
\begin{eqnarray*}
  \mbox{maximize}_{\vy}  & \vb^T \vy \label{dual2} & \\
  \mbox{subject to} & \mA^T\vy + \vz = \vc , & \vz \ge 0.
\end{eqnarray*}
\end{subequations}

Primal optimality is obtained using (\ref{log-bar-der-x-1}) \cite{GHare}
\BE \vy^T \mA = \vc - \mu \mX^{-1}\mathbf{1}. \label{opt-primal} \EE
Substituting (\ref{opt-primal}) in (\ref{dual2}) we get the connection
between the primal and dual
\[ \mu \mX^{-1} \mathbf{1}= \vz. \]
In total, we have a primal-dual system (again we assume that the
solution is strictly feasible, namely $\vx > 0, \vz > 0$)
%\vspace{-5mm}
\begin{subequations}
\begin{eqnarray*}
  \mA\vx = \vb, & \vx > 0, \\
  \mA^T \vy + \vz = \vc, & \vz > 0, \\
  \mX\vz = \mu \mathbf{1}.  &  \\
\end{eqnarray*}
\end{subequations}
The solution $[\vx(\mu), \vy(\mu), \vz(\mu)]$ of these equations
constitutes the central path of solutions to the logarithmic
barrier method \cite[11.2.2]{BV04}. Applying the Newton method to
this system of equations we get \BE \label{newton-3-3-matrix}
\left(%
\begin{array}{ccc}
  0 & \mA^T & I \\
  \mA & 0 & 0 \\
  \mZ & 0 & \mX \\
\end{array}%
\right)
\left(%
\begin{array}{c}
  \Delta \vx \\
  \Delta \vy \\
  \Delta \vz \\
\end{array}%
\right) =
\left(%
\begin{array}{c}
  \vb - \mA\vx \\
  \vc - \mA^T\vy - \vz \\
  \mu \mathbf{1} - \mX\vz \\
\end{array}%
\right). \EE The solution can be computed explicitly by

\BE
\begin{array}{cl}
  \Delta \vy = & (\mA\mZ^{-1}\mX\mA^T)^{-1} \cdot \\
 &  (\mA\mZ^{-1}\mX(\vc - \mu \mX^{-1}\mathbf{1} -\mA^T\vy) + \vb - \mA\vx), \\
  \Delta \vx =& \mX\mZ^{-1}(\mA^T \Delta \vy + \mu \mX^{-1}\mathbf{1} = \vc
+ \mA^T\vy), \\
  \Delta \vz = & -\mA^T\Delta \vy + \vc - \mA^T\vy - \vz.  \\
\end{array}
\nonumber \EE The main computational overhead in this method is the
computation of $(\mA\mZ^{-1}\mX\mA^T)^{-1}$, which is derived from
the Newton step in (\ref{newton-step}).

 Now we would like to use GaBP for computing the solution.
 We make the following simple change to (\ref{newton-3-3-matrix})
 to make it symmetric: since $\vz > 0$, we can multiply the third
 row by $\mZ^{-1}$ and get a modified symmetric system
\[ \label{newton-3-3-matrix-mod}
\left(%
\begin{array}{ccc}
  0 & \mA^T & I \\
  \mA & 0 & 0 \\
  I & 0 & \mZ^{-1}\mX \\
\end{array}%
\right)
\left(%
\begin{array}{c}
  \Delta \vx \\
  \Delta \vy \\
  \Delta \vz \\
\end{array}%
\right) =
\left(%
\begin{array}{c}
  \vb - \mA\vx \\
  \vc - \mA^T\vy - \vz \\
  \mu \mZ^{-1} \mathbf{1} - \mX \\
\end{array}%
\right). \]
Defining $ \tilde{\mA} \triangleq \left(%
\begin{array}{ccc}
  0 & \mA^T & I \\
  \mA & 0 & 0 \\
  I & 0 & \mZ^{-1}\mX \\
\end{array}%
\right),$ and $\tilde{\vb} \triangleq \left(%
\begin{array}{c}
  \vb - \mA\vx \\
  \vc - \mA^T\vy - \vz \\
  \mu \mZ^{-1} \mathbf{1} - \mX \\
\end{array}%
\right). $ one can use GaBP iterative algorithm shown in
Table~\ref{tab_summary}.

In general, by looking at (\ref{taylor-gradient}) we see that the
solution of each Newton step involves inverting the Hessian matrix
$f''(\vx)$. The state-of-the-art approach in practical
implementations of the Newton step is first computing the Hessian
inverse $f''(\vx)^{-1}$ by using a (sparse) decomposition method
like (sparse) Cholesky decomposition, and then multiplying the
result by $f'(\vx)$. In our approach, the GaBP algorithm computes
directly the result $\Delta \vx$, without computing the full
matrix inverse. Furthermore, if the GaBP algorithm converges, the
computation of $\Delta \vx$ is guaranteed to be accurate.

\section{New Convergence Results} \label{convergence} \label{conv-speed}
In this section we give an upper bound on the convergence rate of
the GaBP algorithm. As far as we know this is the first
theoretical result bounding the convergence speed of the GaBP
algorithm.

Our upper bound is based on the work of Weiss \etal~\cite[Claim
4]{Weiss}, which proves the correctness of the mean computation.
Weiss uses the pairwise potentials form\footnote{Weiss assumes
scalar variables with zero means.}, where
\BEA p(\vx) &\propto& \Pi_{i,j} \psi_{ij}(x_i,x_j) \Pi_i \psi_i(x_i), \nonumber \\
 \psi_{i,j}(x_i, x_j) &\equiv& \exp(-1/2 (x_i\  x_j)^T \mV_{ij} (x_i\ x_j)), \nonumber \\
 \mV_{ij} &\equiv& \left(%
\begin{array}{cc}
  \tilde{a}_{ij} & \tilde{b}_{ij} \\
  \tilde{b}_{ji} & \tilde{c}_{ij} \\
\end{array}%
\right). \nonumber \EEA Assuming the optimal solution is $\vx^{*}$, for a
desired accuracy $\epsilon||\vb||_{\infty}$ where
$||\vb||_{\infty} \equiv \max_i |\vb_i|$, and $\vb$ is the shift
vector, we need to run the algorithm for at most $t = \ceil{{
\log(\epsilon)}/{ \log(\beta) }}$ rounds to get an accuracy of
$|x^* - x_t| < \epsilon||\vb||_{\infty}$ where $\beta =
\max_{ij}|\tilde{b}_{ij}/\tilde{c}_{ij}|$.

The problem with applying Weiss' result directly to our model is
that we are working with different parameterizations. We use the
{\em information form} $ p(\vx) \propto \exp(-1/2\vx^T\mA\vx +
\vb^T\vx). $ The decomposition of the matrix $\mA$ into pairwise
potentials is not unique. In order to use Weiss' result, we
propose such a decomposition. Any decomposition from the canonical
form to the pairwise potentials form should be subject to the
following constraints \cite{Weiss} \ \BE \tilde{b}_{ij} = \mA_{ij}
, \ \ \ \ \  \Sigma_j \tilde{c}_{ij} = \mA_{ii}. \nonumber \EE We propose to
initialize the pairwise potentials as following. Assuming the
matrix $\mA$ is diagonally dominant, we define $\varepsilon_i$ to
be the non negative gap \BE \varepsilon_i \triangleq |\mA_{ii}| -
\Sigma_{j}|\mA_{ij}| > 0. \label{varepsilon} \nonumber \EE and the
following decomposition \BE \tilde{b}_{ij} = \mA_{ij} , \ \ \ \
\tilde{c}_{ij} = \mA_{ij} + \varepsilon_i / |N(i)|, \nonumber \EE where
$|N(i)|$ is the number of graph neighbors of node $i$. Following
Weiss, we define $\gamma$ to be \BE \gamma = \max_{i,j}
\frac{|\tilde{b}_{ij}|}{|\tilde{c}_{ij}|} = \frac{|a_{ij}|}{|a_{ij}| +
\varepsilon_i /|N(i)|} = \nonumber \EE \BE = \max_{i,j} \frac{1}{1 + (
\varepsilon_i) /(|a_{ij}||N(i)|)} < 1. \label{gamma} \EE In total, we get
that for a desired accuracy of $\epsilon||\vb||_{\infty}$ we need
to iterate  for $t = \ceil{ {\log(\epsilon)}/{ \log(\gamma)}}$
rounds. Note that this is an upper bound and in practice we indeed
have observed a much faster convergence rate.

The computation of the parameter $\gamma$ can be easily done in a
distributed manner: Each node locally computes $\varepsilon_i$,
and $\gamma_i = \max_{j} {1}/{( 1 + |a_{ij}|\varepsilon_i/N(i)
)}$. Finally, one maximum operation is performed globally, $\gamma
= \max_{i} \gamma_i$.

\subsection{Applications to Interior-Point Methods}
We would like to compare the running time of our proposed method
to the Newton interior-point method, utilizing our new convergence
results of the previous section. As a reference we take the
Karmarkar algorithm~\cite{Karmarkar} which is known to be an
instance of the Newton method \cite{Karmarkar-newton}. Its running
time is composed of $n$ rounds, where on each round one Newton
step is computed. The cost of computing one Newton step on a dense
Hessian matrix is $O(n^{2.5})$, so the total running time is
$O(n^{3.5})$.

Using our approach, the total number of Newton iterations, $n$,
remains the same as in the Karmarkar algorithm. However, we
exploit the special structure of the Hessian matrix, which is both
symmetric and sparse. Assuming that the size of the constraint
matrix $\mA$ is $n \times p, \ \ \ p < n$, each iteration of GaBP
for computing a single Newton step takes $O(np)$, and based on our
new convergence analysis for accuracy $\epsilon||\vb||_{\infty}$
we need to iterate for $r = \ceil{{
\log(\epsilon)}/{\log(\gamma)}}$ rounds, where $\gamma$ is defined
in (\ref{gamma}). The total computational burden for a single
Newton step is $O(np{ \log(\epsilon)}/{\log(\gamma)})$. There are
at most $n$ rounds, hence in total we get $O(n^2p{
\log(\epsilon)}/{\log(\gamma)})$. %relative to $O(n^{3.5})$ of the
%Karmarkar algorithm.

\section{Experimental Results}
\label{exp-results} We demonstrate the applicability of the
proposed algorithm using the following simple linear program
borrowed from \cite{WikiKarmarkar}
\begin{subequations}
\begin{eqnarray*}
  \mbox{ maximize } & x_1 + x_2 & \\
  \mbox{ subject to} & 2px_1 + x_2 \le p^2 +1\,,&\\
  & p=0.0,0.1,\cdots,1.0\,\,. &
\end{eqnarray*}
\end{subequations}

\begin{figure}
\begin{center}
  \includegraphics[scale=0.5, bb=49 234 554 585]{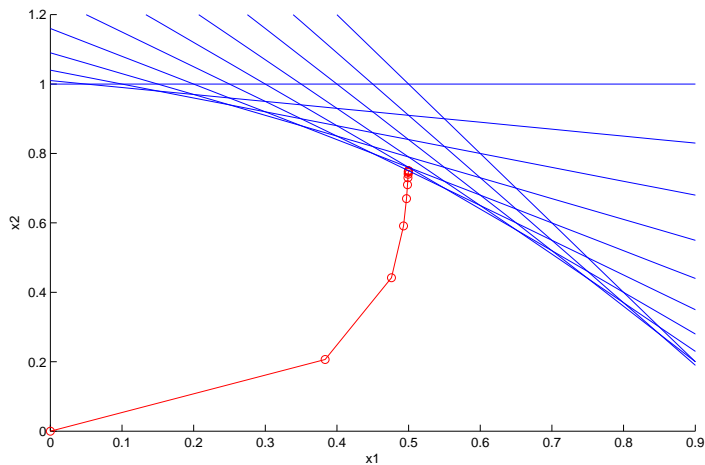}\\
  \caption{A simple example of using GaBP for solving linear programming with two variables and eleven constraints. Each red circle shows one iteration of the Newton method.}\label{toy-example}
\end{center}
\end{figure}

Fig. \ref{toy-example} shows execution of the affine-scaling
algorithm \cite{Affine-scaling}, a variant of Karmarkar's
algorithm \cite{Karmarkar}, on a small problem with two variables
and eleven constraints. Each circle is one Newton step. The
inverted Hessian is computed using the GaBP algorithm, using two
computing nodes. Matlab code for this example can be downloaded
from \cite{ToyNewton}.

Regarding larger scale problems, we have observed rapid
convergence (of a single Newton step computation) on very large
scale problems. For example, \cite{PPNA08} demonstrates
convergence of 5-10 rounds on sparse constraint matrices with
several millions of variables. \cite{ECCS08} shows convergence of
dense constraint matrices of size up to $150,000 \times 150,000$
in 6 rounds, where the algorithm is run in parallel using 1,024
CPUs. Empirical comparison with other iterative algorithms is given
in \cite{Allerton}.

\section{Conclusion}\label{conc} In this paper we have shown how
to efficiently and distributively solve interior-point methods
using an iterative algorithm, the Gaussian belief propagation
algorithm. Unlike previous approaches which use discrete belief
propagation and gradient descent methods, we take a different path
by using continuous belief propagation applied to interior-point
methods. By shifting the Hessian matrix inverse computation
required by the Newton method, from linear algebra domain to the
probabilistic domain, we gain a significant speedup in performance
of the Newton method. We believe there are numerous applications
that can benefit from our new approach.

\section*{Acknowledgement}
O. Shental acknowledges the partial support of the NSF (Grant
CCF-0514859). D. Bickson would like to thank Nati Linial from the
Hebrew University of Jerusalem for proposing this research
direction. The authors are grateful to Jack Wolf and Paul Siegel
from UCSD for useful discussions and for constructive comments on
the manuscript.

\bibliographystyle{IEEEtran}   % (uses file "plain.bst")
\normalsize
\bibliography{IEEEabrv,final_lp}       % expects file "myrefs.bib"

\end{document}